\documentclass[eprint, twocolumn]{revtex4-2}
\usepackage{bbm}
\usepackage{natbib}
\usepackage{graphics}
\usepackage{amsmath}
\usepackage{mathrsfs}
\usepackage{theorem}
\usepackage{float}
\usepackage{hyperref}
\usepackage{rotating}
\usepackage{amssymb}
\usepackage[english]{babel}
\usepackage{color}
\usepackage{fancybox}

\newcommand{\ket}[1]{|#1\rangle}
\newcommand{\bra}[1]{\langle #1|}
\newcommand{\inp}[2]{\langle #1 | #2\rangle}
\newcommand{\be}{\begin{eqnarray}}
\newcommand{\ee}{\end{eqnarray}}
\newtheorem{theorem}{Theorem}

\begin{document}

\title{Effective Hamiltonian approach to the exact dynamics of open system by complex discretization approximation for environment }
\author{H. T. Cui $^{1}$}
\email{cuiht01335@aliyun.com}
\author{Y. A. Yan $^{1}$}
\email{yunan@ldu.edu.cn}
\author{M. Qin $^{1}$}
\email{qinming@ldu.edu.cn}
\author{X. X. Yi $^{2}$}
\email{yixx@nenu.edu.cn}
\affiliation{$^1$ School of Physics and Optoelectronic Engineering \& Institute of Theoretical Physics, Ludong University, Yantai 264025, China}
\affiliation{$^2$ Center for  Quantum Sciences, Northeast Normal University, Changchun 130024, China}
\date{\today}

\begin{abstract}
The discretization approximation method  commonly used to simulate the dynamics of quantum  system coupled to the environment in continuum often suffers from the periodically  partial recovery of initial state because of the effect of finite dimension, dubbed the recurrence. To address this issue, we proposes a  generalization of the discretization approximation method into the complex frequency space basing on complex Gauss quadratures. An effective Hamiltonian can be established by this way, which is  non-Hermitian and demonstrates the complex energy modes with negative imaginary part, describing  the dissipation of the system. This method is applied to examine the dynamics in two exactly solvable models,  the dephasing model and the single-excitation dissipative  dynamics in the Aubry-Andr\'{e}-Harper model. By comparison with the exact numerics and analytical results, it is found that our approach not only significantly reduces the effect of recurrence and improve the effectiveness of calculation, but also provide a unique perspective into the dynamics of open system  from the point of complex energy levels. Furthermore,  we establish a simple relationship between the parameters in computation and the effectiveness of simulation by analyzing the computational error.
\end{abstract}


\maketitle


\section{introduction}

A quantum system is inherently connected to its surrounding environment, and finally becomes  equilibrated. However, rigorous numerical or analytical description of the dynamics of open systems appears to be extremely difficult  as one has to deal with large or infinite number of degrees of freedom in environment. Hence, assumptions such as the weak system-environment coupling and vanishing correlation times of the environment, known as Born-Markov approximation, are usually invoked to find compact and solvable equations \cite{breuer2002}. These approaches sacrifice the accuracy for the sake of operability  and may become  incorrect in some circumstances. Indeed, recent experimental advances have demonstrated the instances  that cannot be formulated under the Born-Markov approximation, such as in solid-state \cite{solid-state} and artificial light-matter systems \cite{lm1, lm2}, as well as in quantum biology\cite{quantbio} and chemistry \cite{quantchem}. In certain situations, nonequilibrium dynamics have also been observed, which demonstrate the importance of the system's dynamics in relation to the non-Markovianity or memory effect of environment \cite{quanttherm1,quanttherm2, memory1, memory2, memory3}. Thus, it is crucial for the research of open dynamics of system to take the property of environment into account completely.

Simulating the open quantum systems imposes a theoretical challenge when the influence of  environment is fully included, often requiring the use of perturbation expansion or numerical simulation  \cite{vega2017}. One common approach is to employ projection operator techniques \cite{breuer2002}, which can yield the Nakjima-Zwanzig equation \cite{nzeqn1, nzeqn2} or the time-convolutionless master equation \cite{tcl1, tcl2, tcl3}. These equations can be used to derive effective master equations through perturbational expansion based on the strength of the system-environment coupling.  On the other hand, several numerical methods have been proposed to determine the exact dynamics of system. For instance, stochastic Schr\"{o}dinger equations (SSE) introduces the stochastic trajectory to simulate the influence of environment \cite{ss1, ss2}. The density matrix of system can be recovered as a sum of projectors of stochastic trajectories. The advantages of  SSE include the preservation of the positivity of the density matrix and the scalability of the wave function with the system basis dimension. Another method involves defining the influence of the environment on the system through a stochastic field.   This leads to the derivation of the hierarchical equations of motion (HEOM) based on the stochastic formulation, allowing for the calculation of the density operator of system including all orders of the system-environment interactions \cite{hem1, hem2}. The path integral methods constitutes a very convenient framework for performing numerical simulations of quantum dynamics and equilibrium quantum statistical mechanics, considering real and imaginary time evolution \cite{pathintegral1, pathintegral2}. Tensor networks techniques adopt a graphical means to represent and reason the dynamics of open quantum\cite{tensornetworks1, tensornetworks2}. They are built on genuine quantum correlations and therefore accurately describe the influence of the environment through the influence functional.

Instead of focusing on the dynamics of system, an alternative method for studying  open quantum systems involves incorporating the complete dynamics of both the system and its environments \cite{discretization1, discretization2}. As for  the  Hamiltonian for environments in continuum can be represented as the integrals, Gaussian quadrature is applied to discretize the integrals, resulting in a chain representation of the total Hamiltonian. Consequently, the dynamics of open quantum system is transformed into the dynamics of isolated many-body system,  which  can be determined exactly by well developed powerful numerical techniques. Moreover, the efficiency and convergence of discretiztion approximation  have been demonstrated  clearly \cite{trivedi21}. A main feature of discretiztion is that  it  not only allows for exact simulation of the dynamics of open system, but also provides the information for effective modes in environment. For the latter, it implies that one can realize the perfect manipulation of system  by finely designed surroundings\cite{muller12}. 

However, the periodic partial recovery of initial state because of the finite dimension of chain Hamiltonian, dubbed recurrence,  can spoil the effectiveness and accuracy of discretization approximation. To tackle this problem, a possible way is to extend the discretization approximation of continuum in environment into the complex frequency domain, resulting in the emergence of complex energy levels that represent dissipation in systems. This method, dubbed complex discretization approximation (CDA),  was initially proposed to examine resonance decay in systems with a discrete state connected to a continuum \cite{bc84, kazansky97, shenvi08}. The idea is to  discretize integrals using Gauss quadrature rules to obtain finite summations of complex items. This results in dynamical equations that can efficiently simulate decay dynamics due to the occurrence of pseudostates with complex energies. However, this way is too specific to apply to more complex situations, where the compact dynamical equations is not always found.

In this paper, the complex discretization approximation (CDA) is reconstructed from a fundamental perspective. The approach involves introducing complex orthonormalized polynomials and applying a unitary transformation to achieve the discretization. The resulting effective full Hamiltonian is non-Hermitian, and thus possess the complex energy modes with negative imaginary part, which describe the dynamics of dissipative system.  It is emphasized that this idea was first used to deal with the dissipative phase transition of the extended Jaynes-Cummings model in our previous work \cite{cui}.  However, a  general approach in this place is proposed so as to tackle the dissipation in more complex systems. The paper is divided into six sections, starting with a general description for the system and its environment in Section \ref{section:model}. Section \ref{section:complexDA} explains briefly how to expand the discretization approximation into the complex frequency domain by introducing the complex orthonormalized polynomials. The method is illustrated for two exactly solvable model in section \ref{exemplification}, the dephasing model and the single-excitation dynamics of open Aubry-Andr\'{e}-Harper model (AAH). The computational error  is discussed explicitly  in section \ref{section:error}, where a simple relation between the parameters in calculation  and the evolution time  can be established. Finally, conclusions and discussion are given at the end of paper.

\section{General description of the open quantum system}\label{section:model}

It is convenient to  model the environment as a set of independent frequency modes, labelled by subscript $k$. Thus, the total Hamiltonian can be written as
\be\label{H}
H&=&H_s + H_b + H_{int} \\
H_b &=& \sum_x \omega_x b_x^{\dagger} b_x;\nonumber \\
H_{\text{int}}&=& \sum_{n, x}\left[g_x F^{\dagger}_n b_x + g^*_x b^{\dagger}_x F_n\right], \nonumber
\ee
in which $b^{\dagger}_x, b_x$  is creation or  annihilation operator for the  $x$-th mode in environment. Supposing the bosonic environment at zero temperature,  the commutative relations
\be
\left[b_x, b^{\dagger}_{x'}\right]=\delta_{x, x'}\nonumber
\ee
are satisfied. $H_s$ is assumed to be expressed in a matrix form, with the subscript $n$ denoting the distinct degree of freedom in the system.  $H_{\text{int}}$ describes the coupling between the system and its environment with the coupling strength $g_x$. $F_n$ or  $F^{\dagger}_n$ represents the physical operator, of which the form is relevant to the specific system. Although $H_{\text{int}}$ is expressed as the form under rotating-wave approximation,  the following derivation does not dependent on this specification. Importantly,  $F_n$  can be Hermitian, for which $H_{\text{int}}$ provides a general description for  the coupling between the system and its environment. 

For a large number of degree of freedom in environment, it is convenient to character the interaction by the spectral function, defined as
\be\label{J}
J\left(\omega\right)= \sum_x \left| g_x\right|^2\delta\left(\omega -\omega_x\right),
\ee
In the limit of continuous $x$, one has the correspondence  $\omega_x\rightarrow \omega(x)$ and  $g_x\rightarrow g(x)$. Thus the spectral function may be written in the continuum as \cite{bulla1997}
\be\label{sd}
J \left[ \omega(x)\right]=  \left| g\left[\omega^{-1}(x)\right]\right|^2\frac{\text{d}x}{\text{d}\omega(x)},
\ee
where $\omega^{-1}(x)$ is the inverse  of  $\omega(x)$ and $\frac{\text{d}x}{\text{d}\omega(x)}$ represents the density of state of environment. In this sense, one can relate the frequency $\omega_k$ to the variable $x$. By this transformatin, $H_b$ and  $H_{\text{int}}$ can be expressed equivalently as
\be\label{integralform}
H_b &\Rightarrow& \int_0^{\infty} \text{d}x \omega(x)b^{\dagger}_x b_x;\nonumber \\
H_{\text{int}}&\Rightarrow& \sum_{n}\int_0^{\infty} \text{d}x \left[g(x)F^{\dagger}_n b_x + g^*(x)b^{\dagger}_x F_n\right],
\ee
As a consequence, $\left[b_x, b^{\dagger}_{x'}\right]= \delta(x-x')$ is a natural requirement.

It is noteworthy that $\omega(x)$ or $g(x)$ are not unique for a given $J\left[ \omega(x)\right]$. By this freedom, one can choose properly the  form of  $\omega(x)$ or $g(x)$ for the sake of  discretization of environment \cite{discretization1}. Given Ohmic spectral function
\be\label{ohmic}
J\left[\omega(x) \right]= \eta \omega_c \left[\frac{\omega(x)}{\omega_c}\right]^{s}e^{-\omega(x)/\omega_c},
\ee
one may choose
\be\label{gx}
\omega(x)=\omega_c x,
g(x)= \sqrt{\eta} \omega_c x^{s/2} e^{- x/2}.
\ee
Assuming that $\omega(x)$ has a linear relationship with dimensionless $x$ is beneficial for the discretization of environment. In \ref{appendix:realGQ}, a brief overview of the discretization approximation method in real frequency space is outlined, and then is applied to simulate the open dynamics in two exactly solvable models. It is illustrated that the simulation is hampered by recurrence, consequently making the calculation  ineffective.

\section{Complex Gauss quadratures and Chain-Mapping in complex frequency space}\label{section:complexDA}

\begin{figure}
\center
\includegraphics[width=5cm]{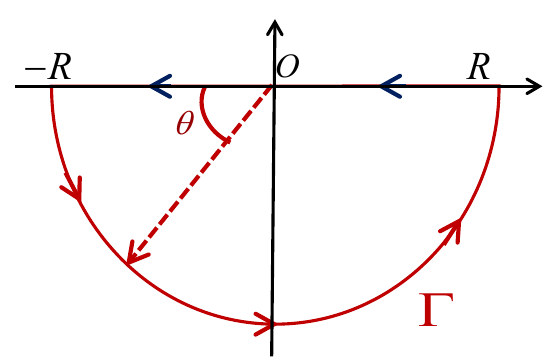}
\caption{(Color online) The diagrammatic sketch of   integral path in the inner product defined by Eq. \eqref{complexinp}. The red solid line, labelled by $\Gamma$,  denotes a semicircle with radius $R$ on the lower half of complex plane. }
\label{fig:contour}
\end{figure}

To reduce the recurrence, we extend the discretization approximation into the complex frequency space in this section. Similar to the procedure shown in \ref{appendix:realGQ}, the first step is to found the  complex  Gauss quadratures.

\subsection{Complex Gauss Quadratures}
The theory of complex orthogonal polynomials can date back to  G. Szeg\"o \cite{szego}, where the unit circle at center $z=0$ is introduced in the complex plane to find a unique sequence of polynomials (called Szeg\"o polynomials). It was known that the zero points of Szeg\"o polynomials are contained in the unit circle. However,  in order to characterize the dissipation, it is necessary to restrict the complex orthogonal polynomials in the lower half of complex plane such that the zero point can displays  negative imaginary part. For this purpose, we adopt the  method presented in Refs. \cite{gautschi86,gautschi87}, by which  a  contour in the lower half of complex plane, as sketched in Fig. \ref{fig:contour}, is chosen accordingly. By this way, one can determine a unique sequence  of complex polynomials displaying zero points with negative imaginary parts. For convenience of discussion, $R=1$ is assumed in this section. Thus, $z=e^{\mathbbm{i}\theta}$ with $\mathbbm{i}=\sqrt{-1}$.

The inner product for complex polynomials $f(z)$ and $g(z)$ is defined as the integral  along contour $\Gamma$ \cite{gautschi86,gautschi87},
\be \label{complexinp}
\left\langle f, g \right \rangle_{\Gamma} = \int_{\Gamma} \frac{\text{d}z}{\mathbbm{i} z} w(z) f(z)g(z) = \int_{\pi}^{2\pi} \text{d}\theta w(e^{\mathbbm{i} \theta}) f(e^{\mathbbm{i} \theta}) g(e^{\mathbbm{i} \theta}), \nonumber  \\
\ee
where $\Gamma$ denotes the integral path as a semicircle with radius $R$ on the lower half of complex plane, as shown by red solid line in Fig. \ref{fig:contour}. For weight function $w(z)$, $\text{Re} \int_{\pi}^{2\pi} \text{d}\theta w(e^{\mathbbm{i} \theta}) \neq 0$ is required  such that  the zero point of the determined polynomial displays nonvanishing imaginary part. It is emphasized  that  the inner product is defined deliberately without complex conjugation. Only by this way,  the three-term recurrence relation can be constructed  since $\left\langle z f, g \right \rangle_{\Gamma}=\left\langle f, zg \right \rangle_{\Gamma}$ is satisfied \cite{gautschi86, gautschi87}. Importantly, we stress that $\left\langle f, f \right \rangle_{\Gamma}$ does not correspond to the norm of $f(z)$  as it may  be complex  or even   zero  by Eq. \eqref{complexinp}. In another point, to construct the  system of orthonormalized polynomials $\left\{\eta_n(z), n=0,1,2, \cdots\right\}$ where $n$ denotes the degree of polynomials,  $\left\langle \eta_n, \eta_n \right \rangle_{\Gamma}\neq 0$ must be required. Fortunately, it has been  demonstrated  in Ref. \cite{gautschi87} that there is the one-to-one correspondence between $\left\{\eta_n(z)\right\}$ and its real counterpart, constructed relative to the inner product $\int_{-1}^{1}\text{d}x w(x)f(x)g(x)$. Thus,  $\left\langle \eta_n, \eta_n \right \rangle_{\Gamma}\neq 0$ is the natural result of this correspondence. Actually, it also provides an alternative way to determine $\eta_n(z)$. Consequently, one can always find the sequence of complex polynomials $\left\{\eta_n(z), n=0,1,2,\cdots \right\}$ by Eq.\eqref{complexinp}, which satisfy the orthonormality
\be\label{complexor}
\left\langle \eta_m, \eta_n \right \rangle_{\Gamma}=\int_{\pi}^{2\pi} \text{d}\theta  w(e^{\mathbbm{i} \theta}) \eta_m(e^{\mathbbm{i} \theta}) \eta_n(e^{\mathbbm{i} \theta})=\delta_{m, n}.
\ee

The recurrence relationship can be derived by the same procedure shown in Appendix \ref{appendix:realGQ}.   As for normalized $\eta_n(z)$, it is convenient to present the recurrence as 
\be \label{complexrecurrence}
\sqrt{\nu_{n+1}} \eta_{n+1}(z)= \left(z - \mu_n\right) \eta_{n}(z) -\sqrt{\nu_{n}}\eta_{n-1}(z),
\ee
where
\be
\mu_n=\left\langle z \eta_n, \eta_n\right\rangle_{\Gamma},
\nu_n=\frac{A^2_{n-1}}{A_n^2},
\ee
$A_n$ denotes the coefficient of  $z^n$ of  $\eta_n(z)$. Resultantly, Eq. \eqref{complexrecurrence} can  be rearranged  in a symmetric matrix form
\be\label{complexM}
&&z \left( \begin{array}{c}
             \eta_0 \\ \eta_1 \\ \vdots \\ \eta_{n-1}
           \end{array}\right)
= M_c
\left( \begin{array}{c}
             \eta_0 \\ \eta_1 \\ \vdots \\ \eta_{n-1}
           \end{array}\right)\nonumber
+ \sqrt{\nu_n}\left( \begin{array}{c}
             0\\ 0 \\ \vdots \\ \eta_{n}\end{array}\right); \\
&&M_c=\left(\begin{array}{cccc}
\mu_0 & \sqrt{\nu_1} & 0 & \cdots \\
\sqrt{\nu_1} &\mu_1 & \ddots & 0 \\
0 & \ddots & \ddots & \sqrt{\nu_{n-1}} \\
0 & \cdots & \sqrt{\nu_{n-1}} & \mu_{n-1}
\end{array}  \right).
\ee
By this matrix form, the zero points $z_i$ and the corresponding  weight $w_i$ for $\eta_n(z)$ can be determined by solving  eigenvalues and corresponding eigenfunctions of $M_c$. Noting that $M_c$ is generally complex, the right and left eigenfunctions have relationship $\ket{i}_R= \left(\ket{i}_L\right)^* (i=0,1,2, \cdots, n-1)$. By the similar method shown in Appendix \ref{appendix:realGQ}, one gets
\be\label{complexwi}
w_i=\left[q^{(i)}_0\right]^2 \int_{\pi}^{2\pi} \text{d}\theta w(e^{\mathbbm{i} \theta}).
\ee
where $q^{(i)}_0$ denotes the first element of $\ket{i}_R$ for zero point $z_i$.
It can be proved  that $z_i$ is confined in the closed region bounded by $\Gamma$ and $x$-axis \cite{wilf}.   With these results, one has the theorem for complex Gauss quadratures  \cite{gautschi86,gautschi87}
\begin{theorem}\label{complexGQ}
   Let $w(z)$ be a weight function defined on  integral path $\Gamma$ in Fig. \ref{fig:contour}. Then the complex polynomials $\eta_n(z)$ can present complex zero points  $z_i$ and corresponding weight $w_i (i= 1, 2, \cdots, n)$, which show the properties (i) All $z_i$ are confined in the closed region illustrated in Fig. \ref{fig:contour}; (ii) $w_i$ is complex; (iii) The equivalence
  \be
  \int_{\Gamma} \frac{\text{d}z}{\mathbbm{i} z} w(z) f(z) \simeq \sum_i w_i f(z_i),
  \ee
  is exactly true for polynomial $f(z)$ of degree $\leq (2n-1)$.
\end{theorem}
The proof of Theorem \ref{complexGQ} is same as Theorem \ref{GQtheorem} in Appendix \ref{appendix:realGQ}, which for instance, can be found in Ref. \cite{wilf}. 

\subsection{Mapping to the chain form }

By Cauchy's integral theorem,  one gets for analytic function $f(z)$
\be
&&\int_{C}\text{d}z f(z)= \int_{\Gamma}\text{d}z f(z)- \int_{-R}^{R}\text{d}x f(x) =0 \nonumber \\
&&\therefore \int_{-R}^{R}\text{d}x f(x)=\int_{\Gamma}\text{d}z f(z).
\ee
where $C$ denotes close path depicted by the arrows in Fig. \ref{fig:contour}.  Thus, one can replace $\int_{-R}^{R}\text{d}x $ by $\int_{\Gamma}\text{d}z $. Correspondingly, $H_b$ and $H_{\text{int}}$ in Eq.\eqref{integralform} can be rewritten as
\be \label{complexH}
H_b&=&\int_{\Gamma} \text{d}z \omega(z)b^{\dagger}_z b_z; \nonumber \\
H_{\text{int}}&=&\sum_n  \int_{\Gamma} \text{d}z  \left[ g(z) F_n^{\dagger} b_z + g(z) F_n b^{\dagger}_z\right],
\ee
in which $b_x^{\dagger}, b_x$ is replaced  by $b_z^{\dagger}, b_z$, and the commutative relation becomes $\left[b_z, b^{\dagger}_{z'}\right]=\delta\left(z-z'\right)$.

Similar to the way adopted in Appendix \ref{appendix:realGQ},  we first introduce the transformation
\be\label{complextransformation}
d_n&=& \int_{\Gamma}\text{d}z \sqrt{\frac{w(z)}{\mathbbm{i} z}} \eta_n(z)  b_z\nonumber\\
d_n^{\ddag}&=& \int_{\Gamma}\text{d}z \sqrt{\frac{w(z)}{\mathbbm{i} z}} \eta_n(z)  b_z^{\dagger}
\ee
and the inverse
\be\label{bz}
b_z&=&\sqrt{\frac{w(z)}{\mathbbm{i} z}} \sum_{n=0}^{N_k-1}\eta_n(z) d_n\nonumber\\
b_z^{\dagger}&=&\sqrt{\frac{w(z)}{\mathbbm{i} z}} \sum_{n=0}^{N_k-1}\eta_n(z) d_n^{\ddag}.
\ee
where $N_k$ denotes the degree of polynomial in the discretization.  It should be stressed that $d_n^{\ddag}$ is not the Hermitian conjugation of $d_n$, even though they are related to the annihilation or creation operator $b_z, b^{\dagger}_z$ respectively.  This odd feature is a consequence of the unique definition for inner product Eq. \eqref{complexinp}.  By Eq. \eqref{complexor}, it is easy to find
\be
\left[d_n, d_m^{\ddag}\right]&=&\iint \text{d}z \text{d}z' \sqrt{\frac{w(z)w(z')}{\mathbbm{i}^2 z z'}} \eta_m(z) \eta_n(z') \left[b_z, b_{z'}^{\dagger}\right]\nonumber \\
&=&\int\text{d}z \frac{w(z)}{\mathbbm{i} z}\eta_m(z) \eta_n(z)=\delta_{m, n}.
\ee
Specially, it should be pointed out the transformation in Eqs. \eqref{bz} is unitary only if $N_k$ is infinite. This effect of finite $N_k$ is responsible for the computational errors.

Substituting Eqs. \eqref{bz} into Eqs.\ref{complexH}, one obtains for $H_b$
\be
H_b= \omega_c \sum_{i, j}d_i^{\ddag} d_j  \int_{\Gamma}\text{d}z \frac{w(z)}{\mathbbm{i} z} z \eta_i(z) \eta_j(z).
\ee
Replacing $z \eta_i(z) $ by rearranging Eq. \eqref{complexrecurrence} and using  Eq. \eqref{complexor}, one gets
\be\label{hb}
H_b&=& \omega_c \left(\cdots,  d_m^{\ddagger}, \cdots\right)M_c
\left( \begin{array}{c}\vdots \\ d_n \\ \vdots \end{array}\right),
\ee
in which $M_c$ is defined by Eq. \eqref{complexM}. Apparently, the eigenvalues  of $M_c$ correspond  to the discrete modes  in environment, which is complex due to the non Hermiticity of $M_c$. Defining  the new mode operator
\be \label{tildeD}
\widetilde{d}_i&=& \sqrt{w_i} \sum_{n=0}^{N_k-1} \eta_n(z_i) d_n,\nonumber \\
\widetilde{d}_i^{\ddagger}&=& \sqrt{w_i} \sum_{n=0}^{N_k-1} \eta_n(z_i) d_n^{\ddagger},
\ee
$H_b$ then can be diagonalized as
\be
H_b =\omega_c\sum_{i=0}^{N_k-1} z_i \widetilde{d}_i^{\ddagger}\widetilde{d}_i.
\ee

As for $H_{\text{int}}$, one gets
\be\label{complexhint}
H_{\text{int}}= \sum_{n=1}^{N_s}\sum_{i=0}^{N_k-1} \int_{\Gamma}\text{d}z \sqrt{\frac{w(z)}{\mathbbm{i} z}} \eta_i(z)\left[g(z) F^{\dagger}_n d_i +  g^* (z)d_i^{\ddagger} F_n \right],\nonumber \\
\ee

\subsection{Simplification of  $H_{\text{int}}$ }
To simplify Eq. \eqref{complexhint} further, one has to deal with the integration. It appears that $\int_{\Gamma}\text{d}z \sqrt{\frac{w(z)}{\mathbbm{i} z}} \eta_i(z) g(z)$ could be evaluated numerically only if $\eta_i(z)$ is determined. However, our empirical analysis has revealed that this methodology is not feasible. Instead, the right way is to approximate the integration according to  Theorem \eqref{complexGQ}, 
\be
&&\int_{\Gamma}\text{d}z \sqrt{\frac{w(z)}{\mathbbm{i} z}} \eta_i(z) g(z) = \int_{\Gamma}\text{d}z \frac{w(z)}{\mathbbm{i} z} \sqrt{\frac{\mathbbm{i} z}{w(z)}} \eta_i(z) g(z) \nonumber \\
&&\simeq\sum_{j=0}^{N_k-1} \sqrt{\frac{\mathbbm{i} z_j}{w(z_j)}} w_j \eta_i(z_j) g(z_j).
\ee
Then, Substituting  the expression above into Eq. \eqref{complexhint} and Using Eq. \eqref{tildeD},   one gets 
\be\label{integral}
&&\sum_{n=1}^{N_s}\sum_{i, j=0}^{N_k-1} \sqrt{\frac{\mathbbm{i} z_j}{w(z_j)}} w_j \eta_i(z_j)  g(z_j) F^{\dagger}_n d_i \nonumber \\
& = &\sum_{n}^{N_s}\sum_{j=0}^{N_k-1} \sqrt{\frac{\mathbbm{i} z_j}{w(z_j)}} \sqrt{w_j}g(z_j) F^{\dagger}_n \widetilde{d}_j
\ee

However, the evaluation encounters discrepancy  for the integral  $\int_{\Gamma}\text{d}z \sqrt{\frac{w(z)}{\mathbbm{i} z}} \eta_i(z) g^*(z)$. Our actual calculation shows that the direct simplification of this integral would lead to an inaccurate description of $H_{\text{int}}$, which ravages the efficiency and precision of simulation. The reason for this discrepancy stems from the unique definition of inner product for complex polynomials, see Eq.\eqref{complexinp}, where the complex conjugation is absent for the purpose to construct the three-term recurrence relation. This results in the environmental  operators  $d_n, d^{\ddagger}_n$, defined by Eq. \eqref{complextransformation}, not being complex conjugated to each other, even though they are related to annihilation or creation operator $b_z, b^{\dagger}_z$ respectively.  So, we enforce the following approximation for $H_{\text{int}}$ 
\be \label{improvedhint}
H_{\text{int}}\simeq \sum_{n=1}^{N_s}\sum_{j=0}^{N_k-1} \left(\mathbbm{g}_j F^{\dagger}_n \widetilde{d}_j + \mathbbm{g}^*_j F_n \widetilde{d}^{\ddagger}_j \right)
\ee
where $\mathbbm{g}_j= \sqrt{\frac{\mathbbm{i} z_j}{w(z_j)}} \sqrt{w_j}g(z_j)$. Evidently, the coefficient of second term is replaced by the complex conjugation of that in the first term. The validity of approximation  is demonstrated in the following illustrations, compared to the exact or analytical results.

\subsection{Evaluation of  the full dynamics}

In the case of an Ohmic environment, where $x$ falls within the interval $\left[0, \infty\right)$, it is necessary to perform a translation of $z_j$ to $R(1+z_j)$. This ensures that the real part of $R(1+z_j)$ corresponds to the value of $x$.  Under this translation, one has
\be\label{replacement}
z_j  & \rightarrow &R(1+z_j)\nonumber \\
\mathbbm{g}_j & \rightarrow & \sqrt{\frac{\mathbbm{i} R z_j}{w\left[R(1+z_j)\right]}} \sqrt{w_j} g\left[R(1+z_j)\right],
\ee
The $\mathbbm{i} R z_j$  is a result of the requirement that $\int_{\Gamma} \frac{\text{d}z}{\mathbbm{i} z} =\int_{\pi}^{2\pi}\text{d}\theta$ must be preserved under the translation. The precision of numerical evaluation is strongly relavent to the  value of $R$ since it determines the upper bound of integration.

Consequently, a non-Hermitian effective Hamiltonian, labelled as $H_{\text{eff}}$,  can be obtained to simulate  the full dynamics of system plus its environment.  The eigenfunctions of $H_{\text{eff}}$ satisfy the biorthonormality \cite{brody14}
\be\label{deltamn}
{_L\inp{m}{n}_R}=\delta_{m,n},
\ee
where $\ket{n}_R$ denotes the right eigenfunction of  $H_{\text{eff}}$ with eigenvalue $E_n$, and $\ket{n}_L$ denotes the left eigenfunction with eigenvalue $E^*_n$. By solving the Schr\"{o}dinger equation,  one gets  $\ket{\psi(t)}$ for the system and its environment at any time $t$
\be\label{psit}
\ket{\psi(t)}= \sum_n e^{-\mathbbm{i}E_n t} \ket{n}_R {_L\inp{n}{\psi(0)}}.
\ee

Formally, one can define the evolution operator $U(t)$ and its inverse $U^{-1}(t)$
\be 
U(t)=  \sum_n e^{-\mathbbm{i}E_n t} \ket{n}_R {_L\bra{n}}, 
U^{-1}(t)=U(-t),  
\ee
where $ U^{-1}(t)$ is decided by using the biorthonormality Eq. \eqref{deltamn} and the completeness $\sum_n  \ket{n}_R {_L\bra{n}} = I$.  Thus, one gets
\be
\ket{\psi(t)}= U(t) \ket{\psi(0)}.
\ee
It is stressed that $U(t)$ is nonunitary since $E_n$ can be complex. As a consequence, the bra with respect to $\ket{\psi(t)}$ can be defined as 
\be 
\bra{\overline{\psi(t)}}= \bra{\psi(0)} U^{-1}(t),
\ee
that is a result of the nonHermitianity of $H_{\text{eff}}$. It can be shown readily 
\be
\inp{\overline{\psi(t)}}{\psi(t)}= \inp{\psi(0)}{\psi(0)}=1.
\ee
In this sense, one can define the density operator 
\be 
\rho(t)=\ket{\psi(t)}\bra{\overline{\psi(t)}}=U(t)\rho(0) U^{-1}(t),
\ee
where $\rho(0)=\ket{\psi(0)} \bra{\psi(0)}$. In general,  $\rho(t)$ is nonHermitian since $U(t)$ is nonunitary. Mathematically, $\rho(t)$ is connected to $\rho(a)$ by a similarity transformation, by which the trace is preserved \cite{bgs25}.  As for open quantum system, the density operator of system $\rho_s(t)$ can still be defined by tracing out the degree of freedom of discretized environment,
\be 
\rho_s(t)&=&  \sum_p {_L \inp{p}{\psi(t)}}\inp{\overline{\psi(t)}}{p}_R \nonumber \\
&=& \sum_p {_L \bra{p} U(t) \ket{\psi(0)}} \bra{\psi(0)} U^{-1}(t) \ket{p}_R \nonumber \\
&=& \sum_{p, m,n} {_L\inp{p}{m}_R}  {_L\inp{n}{p}_R} e^{\mathbbm{i}\left(E_n- E_m \right)t} {_L\inp{m}{\psi(0)}} \inp{\psi(0)}{n}_R\nonumber \\
\ee 
in which  $\ket{p}_R, {_L\bra{p}} $ denote the right and left eigenfunction of Eq. \eqref{hb} respectively. Evidently,  $\rho_s(t)$ is trace-preserving, but is not Hermitian.  A recent research shows that because of the nonHermitianity of system, the established properties in Hermitian system relevant to wave function or density operator, e.g. inner product or expectational value of observable,  have to redefined carefully \cite{bgs25}. This question is beyond the scope of this paper. However, it should be stressed that the calculation below is irrelevant to  $\rho(t)$ or $\rho_s(t)$.

\subsection{Simulating the stable or stretched  dynamics of system}
The effective Hamiltonian  $H_{\text{eff}}$ is expected to accurately model the dissipative dynamics of the system since $E_n$ possesses the negative imaginary part, causing the amplitude in Eq. \eqref{psit} to decay exponentially. In another point, the strong coupling between system and environment may lead to the nonequilibrium behavior, such as the stable oscillation between  energy states  or reaching a steady state apart from the equilibrium \cite{lm1, lm2}. To capture these unique dynamics of system more accurately, the improvement for the previous  evaluation is required.

After the explicit calculation, it is observed that replacing the expected $E_n$ with its real part in Eq. \eqref{psit} can result in an excellent simulation for the stable dynamics of system, i.e., utilizing
\be \label{stablepsit}
\ket{\psi(t)}= \sum_n e^{-\mathbbm{i} \text{Re}(E_n) t} \ket{n}_R {_L\inp{n}{\psi(0)}}.
\ee
The effectiveness of this method is examined, compared with the approach by Eq. \eqref{psit} and the exact numerics, as well as the analytical approach shown in the next section. Admittedly, this proposition is empirical since it is aimed to recover the coherent evolution of system only by eliminating the exponential decaying of the amplitude. Regarding the slow-decaying dynamics of system presented in the following section, our calculation shows that while Eq.\eqref{stablepsit} can provide the accurate simulation of the dynamics for a short-term evolution, it appears to be inadequate for  predicating the long-term behavior.

With these preparations, we are ready to simulate the dynamics in the dephasing model and the dissipative AAH model in the single-excitation subspace, as two exemplifications. It will be shown that depending on the status of system, the dynamics of system can displays the decaying, the stable oscillation or the slow-decaying behavior, all of which can be captured correctly by the method proposed in this section.

\begin{figure}[tb]
\center
\includegraphics[width=6cm]{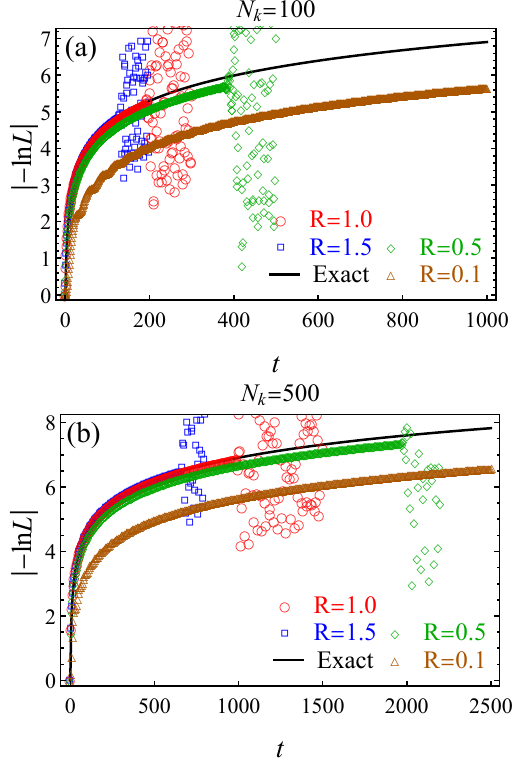}
\caption{(Color online) (a), (b): The plots of the modulus of $- \ln L$ defined in Eq. \eqref{complexL} for different $R$  and  $N_k$. $\eta= \omega_c =1$ is chosen for the plotting. The solid black line is obtained by Eq. \eqref{L}.  The evolution time $t$ is in units of $1/\omega_c$.}
\label{fig:complexL}
\end{figure}

\section{Exemplifications}\label{exemplification}

To ensure effective simulation, it is essential to choose $w(z)$ properly. The explicit calculation  indicates that the choice of $w(z)/\mathbbm{i} z= \eta \omega^2_c z^s e^{-z}$ is not suitable, since it leads to a real symmetric  $M_c$. Resultantly, the eigenvalues of $H_{\text{eff}}$ are real, which cannot improve the simulation. Instead, we choose $w(z)=1$ in the following discussion. As shown in the following illustrates,  the simulation can be significantly improved, in contrast to the calculation in Appendix \ref{appendix:realGQ},

\subsection{Model I: the Dephasing model }

The total hamiltonian is written as \cite{breuer2002}
\be\label{hdephasing}
H=\frac{\omega_0}{2}\sigma_z + \sum_k \omega_k b_k^{\dagger}b_k + \frac{\sigma_z}{2} \sum_k\left(g_k b_k^{\dagger} + g^*_k b_k\right).
\ee
in which $\sigma_z$ denotes the  $z$ component of  Pauli operator, and $b_k^{\dagger}, b_k$ are creation or annihilation operator of  the $k$-th mode in the bosonic environment. It was known  that the decoherence factor $L(t)$ at zero temperature is \cite{breuer2002}
\be \label{lnL}
-\ln L=  \sum_k \left|g_k\right|^2\tfrac{1- \cos \omega_k t}{\omega_k^2} = \int_{0}^{\infty} \text{d}x J(x) \tfrac{1-\cos x t}{x^2}
\ee
Choosing Ohmic spectral density Eq. \eqref{ohmic} and setting  $\eta=\omega_c=1$, one obtains  for $s=1$
\be \label{L}
- \ln L= \frac{1}{2} \ln \left(1 + t^2\right).
\ee

As for the complex discretization approximation, one may replace $\omega_k$ and $g_k$ by $z_j$ and $\mathbbm{g}_j$ respectively. Formally,  $L$ is defined at zero temperature  as  
\be 
L= \left| \bra{\uparrow}\bra{0} U^{-1}(t)  U(t) \ket{0} \ket{\downarrow} \right|
\ee 
where $\sigma_z\ket{\uparrow(\downarrow)}=\pm \ket{\uparrow(\downarrow)}$. Finally, one has
\be \label{complexL}
- \ln L =  & & \left|\sum_j  \left(\sqrt{\frac{\mathbbm{i} R z_j}{w\left[R(1+z_j)\right]}} \sqrt{w_j} g\left[R(1+z_j)\right] \right)^2  \right.\nonumber\\
&& \left .\times \frac{1- \cos R(1+z_j) t}{R^2(1+z_j)^2} \right|,
\ee
for which Eq. \eqref{replacement} is applied. Since $- \ln L$ becomes complex after this transformation, the modulus of $- \ln L$ is illustrated in Fig. \ref{fig:complexL}. Evidently, the effectiveness of simulation are significantly relevant to both $N_k$ and $R$. While the period of time for which $- \ln L$ can be evaluated rigorously is enlarged by increasing $N_k$, the accuracy of computation is enhanced only by increasing $R$. The error analysis will be presented in the next section.

\begin{figure*}[tb]
\center
\includegraphics[width=18cm]{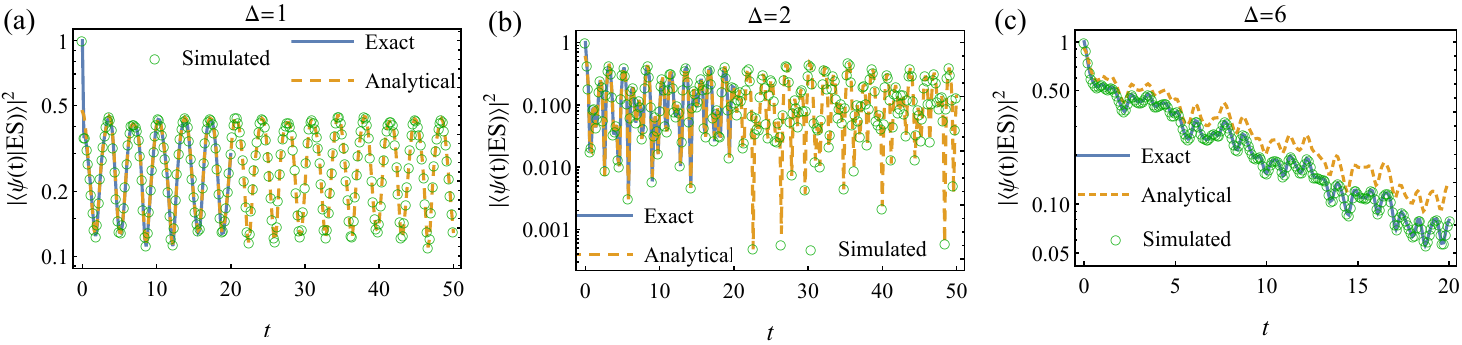}
\caption{(Color online)(a)-(c): The  plot of $\left|\inp{\psi(t)}{ES}\right|^2$ for $\Delta=1, 2$ and $6$. To demonstrate the effectiveness of the simulation (green empty circle), both the exact numerics (solid-blue line) and the analytical approach (dashed-orange line) are presented at the same time.  For the simulation, $N_k=2000$ and $R=4.72$ are chosen, based on the error analysis shown in the next section.  For all plots, we have chosen $N_s=8, \beta=\left(\sqrt{5}-1\right)/2, \phi=\pi$ and $\eta=0.1, \omega_c=10$. Especially, whereas the simulation for  $\Delta=1,2$ is attained by Eq. \eqref{stablepsit},  we adopt Eq. \eqref{psit} for  $\Delta=6$ to find the  compliance with the exact numerics. The evolution time $t$ is in units of $1/J$. } \label{fig:aah}
\end{figure*}

\begin{figure*}[tb]
\center
\includegraphics[width=18cm]{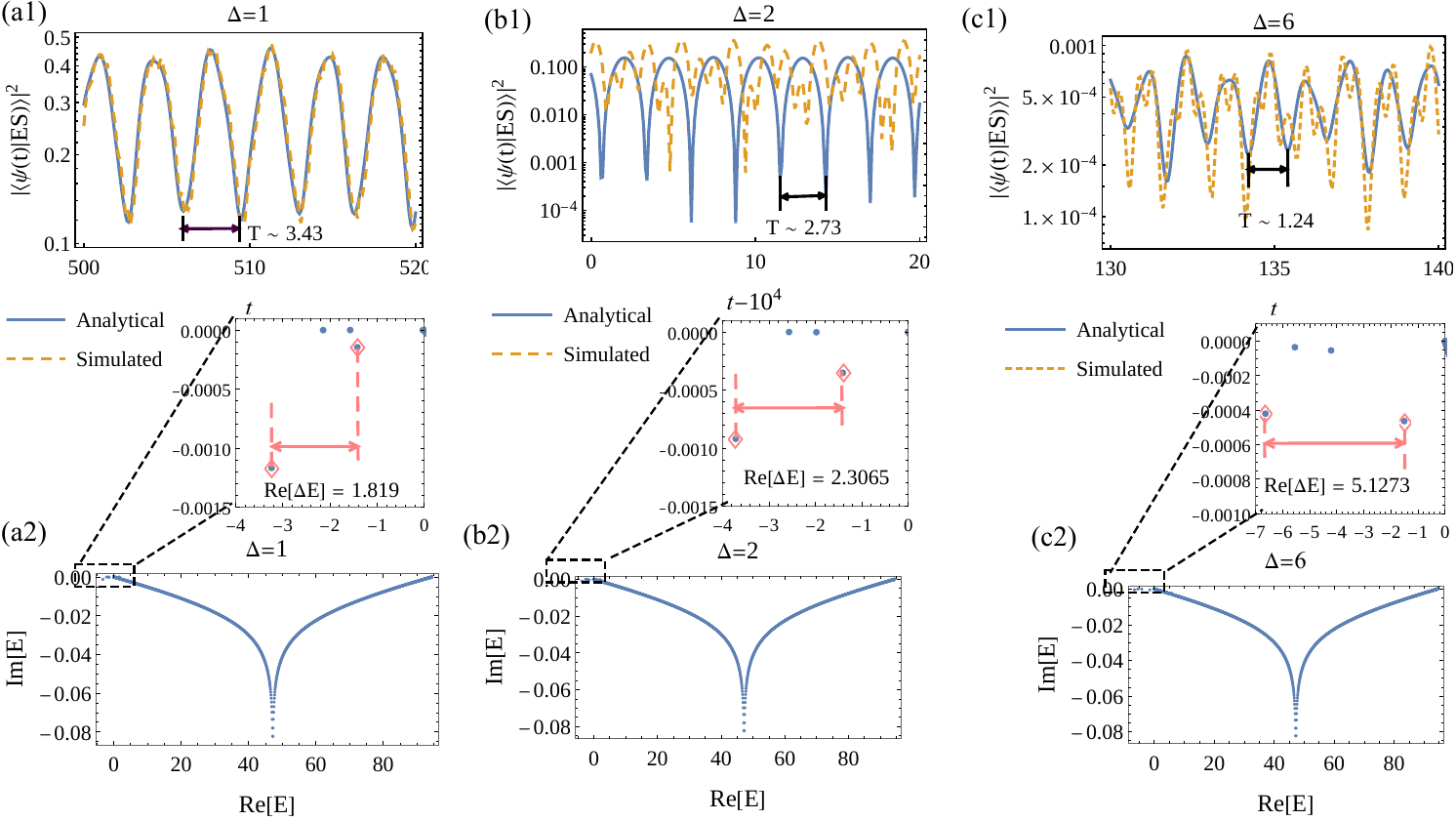}
\caption{(Color online) (a1), (b1), (c1): The  long term behavior for $\left|\inp{\psi(t)}{ES}\right|^2$ when  $\Delta=1, 2$ and $6$.  The evaluation of CDA simulation is shown by the dashed-orange line, while the analytical approach is shown by the solid-blue line. The value of $T$ denotes the average period of oscillation of SP. The simulation of SP is implemented by Eq. \eqref{stablepsit} for $\Delta=1, 2$, while it is done by Eq. \eqref{psit} for $\Delta=6$.  Evolution time $t$ is in units of $1/J$. (a2), (b2), (c2): The complex eigenvalue  $E$ (in units of $J$) of $H_{\text{eff}}$ is obtained  by solving the equation $H_{\text{eff}} \ket{n}_R = E_n \ket{n}_R$.  The value of $\text{Re}\left[ \Delta E \right]$ gives the difference between the two discrete energy levels, which are relevant to the observed oscillation illustrated in (a1), (b1), (c1). The other parameters are the same as in  Fig. \ref{fig:aah}.  } \label{fig:aahlongterm}
\end{figure*}

\subsection{Model II: the Singel-excitation dynamics in AAH model}

Another exactly solvable situation is the dissipative dynamics in the single-excitation subspace. For concreteness, we explore the single-excitation  dynamics in the Aubry-Andr\'{e}-Harper model (AAH) coupled to an Ohmic environment in this subsection. AAH model has recently been discussed intensively since the localization-delocalization transition can be observed experimentally \cite{AABS19}. In this place, we examine its dissipative dynamics by the complex discretization approximation, as well as the exact numerics and the analytical evaluation as comparison.

The Hamiltonian for AAH model is written as \cite{aa80, haper55}
\be\label{AAH}
H_s&=& J \sum_{n=1}^{N_s} \left(c^{\dagger}_{n} c_{n+1} +c^{\dagger}_{n+1} c_{n} \right) + \nonumber \\
&&  + \Delta \sum_{n=1}^{N_s} \cos(2\pi \beta n +\phi) c^{\dagger}_n c_{n},
\ee
where $\Delta$ characterizes the strength of the onsite potential, and $J\equiv 1$ is assumed for brevity. $c_n \left( c^{\dagger}_n\right)$ denotes the annihilation (creation) operator of excitations on the $n$-th lattice site. To avoid the effect of boundary,  the periodic boundary condition$c_{N_s+1}=c_1$ is imposed in the following calculation. It was known that  the  critical point  $\Delta=2$ can separate the extended phase ($\Delta<2$)  from the localized phase ($\Delta>2$) in this model. The recent studies have showed that the localized phase would  be destroyed because of the coupling to environment \cite{levi16, fischer16, luschen17}.

In the single-excitation subspace,  the state of system can be formulated  as
\be\label{state}
\ket{\psi(t)}&=& \left(\sum_{n=1}^{N_s} a_n(t) \ket{1}_n \right) \ket{0}^{\otimes N_k} + \nonumber \\ &&\ket{0}^{\otimes N_s}  \left(\sum_{k} b_k(t) \ket{1}_k\right),
\ee
where $\ket{1}_n = c_n^{\dagger}\ket{0}$, $\ket{0}$ is the vacuum state of environment  and $\ket{1}_k=b_k^{\dagger}\ket{0}$. Substituting Eq. \eqref{state} into Schr\"{o}dinger equation and eliminating $b_k (t)$, one gets
\be\label{evolution}
\mathbbm{i}\frac{\partial }{\partial t}a_n(t)&=& \left[a_{n+1}(t) + a_{n-1}(t)\right]+ \Delta \cos(2\pi \beta n +\phi)a_n(t) \nonumber\\
&-& \mathbbm{i} \sum_{n=1}^{N_s} \int_0^t \text{d}\tau a_n(\tau)\int_0^{\infty} \text{d} \omega J(\omega)e^{-\mathbbm{i} \omega(t-\tau)}.
\ee
Eq. \eqref{evolution} can be solved exactly  by numerical iteration. However, due to the integrals involving time $t$,  the iteration becomes so exhaustive for a long-time evolution. So, the exact numerics is restricted to $t\leq 20 $.

By Laplace transformation, Eq. \eqref{evolution} can be transformed as
\be \label{Ap}
\mathbbm{i} \left[p A_n(p)- a_n(0) \right]&&= \Delta \cos\left(2\pi\beta n + \phi \right) A_n(p)+A_{n+1}(p)+ \nonumber \\
&&  A_{n-1}(p)   - \sum_m A_m(p) \int_{0}^{\infty} \text{d}\omega \frac{J(\omega) }{ \omega- \mathbbm{i} p},
\ee
where  $A_n(p)= \int_{0}^{\infty} \text{d}t a_n(t) e^{- p t}$. In principle, $A_n(p)$ can be decided by solving the system of equation above only if the integral can be evaluated. Thus, $a(t)$ can be determined by inverse Laplace transformation. In Appendix \ref{analyticalsolution}, two distinct types of  solutions for Eq. \eqref{Ap} has been determined.  Thus, the analytical expression for $a_n(t)$ can be obtained, which provides the proper description for the mediated and long-term behavior of the system.

To demonstrate the effectiveness of simulation,  we simulate the single-excitation dissipative  dynamics of AAH model by calculating the modulus of overlap  $\left|\inp{\psi(t)}{ES}\right|^2$, in which the initial state $\ket{ES}$ represents the total state of system and discretized environment with the highest  energy level in AAH model  occupied by single excitation and the environment left in vacuum. Actually, $\left|\inp{\psi(t)}{ES}\right|^2$ gives the survival probability (SP) for single excitation being preserved in highest excited energy level of AAH model. SP is plotted for $\Delta=1, 2$ and $6$ respectively in  Figs. \ref{fig:aah}, for which AAH model can display the extended, critical and localized properties respectively.  Additionally, the analytical and  exact numerical evaluation are also provided  to validate the simulation.  The results show that the simulation aligns perfectly with the analytical and  exact numerical approaches.

There are some comments on the performance of simulation sketched by Fig. \ref{fig:aah}. First, as illustrated in the following section, the choice of $R=4.72$ for $N_k= 2000$ is optimal. For this choice, the simulation is highly effective and reasonably accurate. Secondly, to simulate the stable oscillation illustrated in Fig. \ref{fig:aah}(a),  $\ket{\psi(t)}$ is evaluated using Eq. \eqref{stablepsit}. The validity  of this choice is approved  by observing the consistence between CDA simulation and analytical approach. Moreover, it is found that CDA can provide the reliable simulation of SP even after a long term evolution,  sketched in Fig. \ref{fig:aahlongterm} (a1). This picture means that Eq. \eqref{stablepsit} can provide the correct description for the stability of system.

At the critical point $\Delta=2$,  the evaluation shows that SP has a very slow decaying. Given that AAH model is neither localized nor extended at critical point, this picture would be a manifestation of this uncertainty. It is pointed out that  Eq. \eqref{stablepsit} seems fail to simulate the long term behavior of SP, as shown in  Fig. \ref{fig:aahlongterm} (b1), although it provides the perfect simulation of SP for a short term as shown in Fig. \ref{fig:aah}(b).  

For $\Delta=6$, it is observed that Eq. \eqref{psit} is more suitable for the simulation of SP,  rather than Eq. \eqref{stablepsit}. This is likely due to the exponential decaying of SP observed in  Fig. \ref{fig:aah}(c), which can be captured accurately by the negative imaginary part in the $E$.  Since the analytical evaluation for the short term evolutions becomes less accurate in this case, the exact numerics is employed up to $t=20$.  As shown in Fig. \ref{fig:aah}(c), the  simulation closely matches the numerical results. However, for a larger value of $t$, the computation becomes more exhaustive.  
By a comparative illustration in Fig. \ref{fig:aahlongterm} (c1),  it is found that Eq. \eqref{psit} can still present a reasonable predication of SP for a large $t$.

The eigenvalue $E$ obtained by solving $H_{\text{eff}} \ket{n}_R = E_n \ket{n}_R$,  is studied to understand  the evolution of SP.  As shown by panels (a2), (b2) and (c2) in Figs. \ref{fig:aahlongterm}, the levels of $H_{\text{eff}} $ demonstrates a  band structure, with  multiple discrete energy levels presented in the region $E<0$. These discrete levels are closely related to the single particle bound states, as their real parts  match the energy of the bound state depicted in Fig. \ref{fig:bs}(a1)-(a3). This relationship is  supported further by determining the angular frequency for  oscillation in the evolution of SP, which is entirely decided by single-particle bound states.  As shown in Figs. \ref{fig:aahlongterm} (a2), (b2) and (c2),  the  calculated angular frequency is shown to match the energy difference between two energy levels in the $E<0$ region. This picture implies that the stable oscillation of SP is a result of the transition of single excitation between the two complex levels. We also evaluate  the overlaps between the two related levels and $\ket{ES}$. For $\Delta=1$, the square absolute values of overlaps are $\sim 0.505$ and  $\sim 0.1545$ respectively. In contrast, the overlaps with   the other complex levels are no more than $\sim 10^{-3}$. Similar observation can be found for $\Delta=2$, for which the overlaps are $\sim 0.1934$ and $0.2002$ respectively. As for $\Delta=6$, our calculation shows that the maximal overlap happens for the complex levels embedded in the band. It thus implies that  the information of initial state  would be diluted  by the bath, and may lead to the decay of SP. 

In summary, our method  provides the excellent  simulation for the dissipative dynamics of AAH model.  However, it is necessary to use different evolutions, such as  Eq. \eqref{psit} or Eq. \eqref{stablepsit}, in order to effectively capture the unique  dynamics of system. This may  be due to the different  physical characteristics associated with the stable oscillation, slow-decaying dynamics and exponential decaying. However, the recent studies has revealed that the open quantum system can undergo the dissipative  phase transition, leading to the  intrinsic  changes in the  dynamics \cite{DPS}. The presence of three types of dissipation could be considered as a reflection  of dissipative  phase transition. Therefore, it is not surprising that using only one method cannot  capture all three distinct dynamics of system.

\begin{figure*}[tb]
\center
\includegraphics[width=14cm]{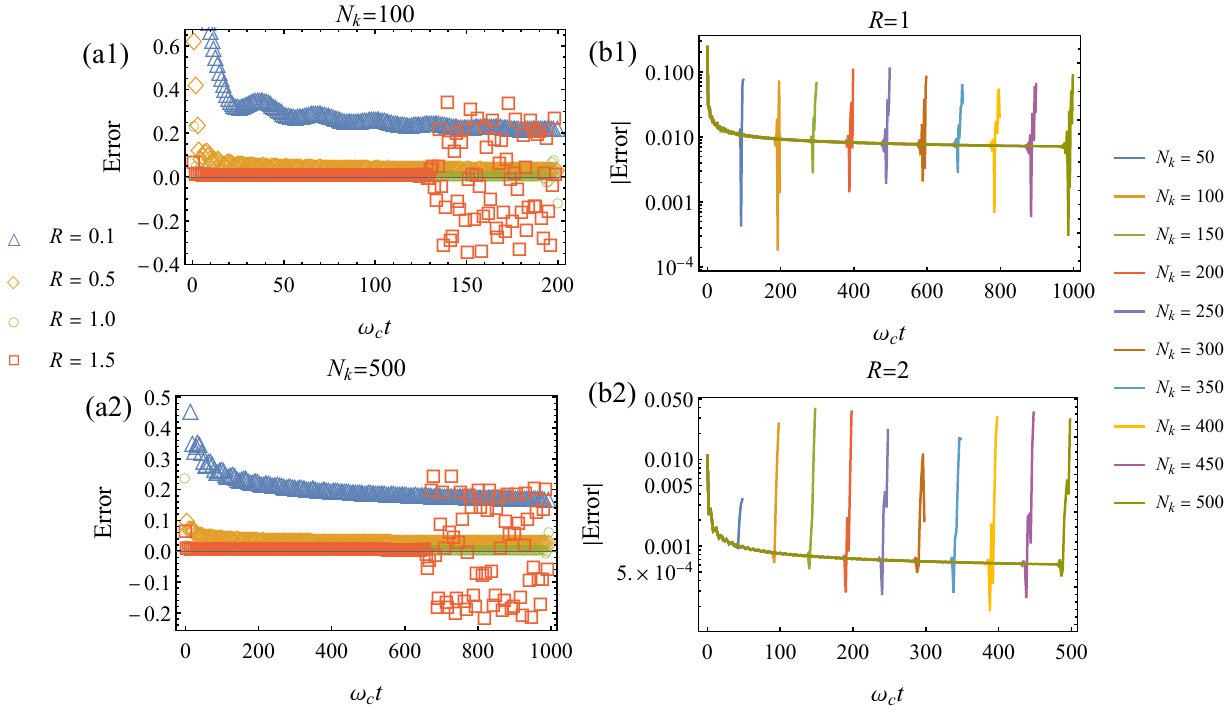}
\caption{(Color online) (a1), (a2): The computational error defined by Eq. \eqref{errorE} for the evaluation of $-\ln L$ is plotted in panels (a1) and (a2) for different $R$. (b1), (b2): the modulus of the Error is plotted in logarithm  for different values of $N_k$. All plots use the same parameters as those in Fig. \ref{fig:complexL}.  }
\label{fig:errordephasing}
\end{figure*}

\begin{figure*}[tb]
\center
\includegraphics[width=17cm]{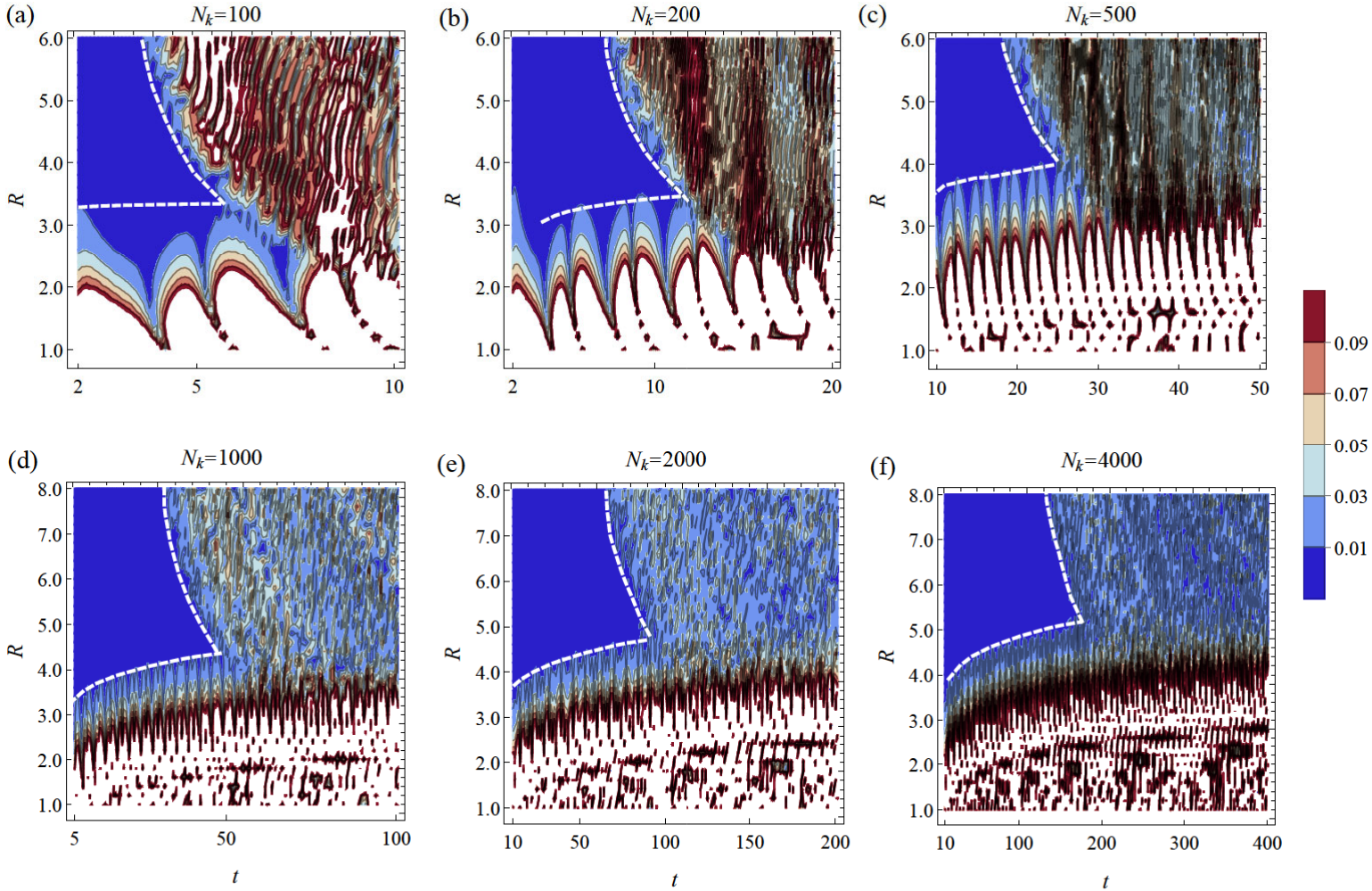}
\caption{(Color online) (a)-(f): The contour plot  for the modulus of Deviation defined by Eq. \eqref{errorA} when $\Delta=1$  versus $R$ and $N_k$. The other parameters are chosen as the same in Fig. \ref{fig:aah}.  The dashed white line characterizes the boundary of the region, where the Deviation is less than 0.01. In the white regions, the modulus of Deviation shows a value larger than 0.1, which is useless for the determination of $t_p$ and thus is not plotted explicitly.   the For the plots, the simulation of SP is reached by Eq. \ref{stablepsit}. The evolution time $t$ is in units of $1/J$.}
\label{fig:errordelta1}
\end{figure*}

\section{ analysis  for the accuracy of simulation }\label{section:error}

To quantify the accuracy of the simulation, two quantities are  elaborated,  which defined respectively as
\be\label{errorE}
\text{Error}&=& \frac{\text{Simulated}- \text{Exact}}{ \text{Exact}}, \\
\label{errorA}
\text{Deviation}&=& \frac{\text{Simulated}- \text{Analytical}}{\text{Simulated}+ \text{Analytical}},
\ee
where ``Simulated", ``Analytical" and ``Exact" denote evaluated SP respectively by the simulation, analytical and exact  methods. The Error depicts the computational error when the exact result can be found.  In contrast, we adopt the Deviation to quantify the difference between the simulation and analytical approach when there is no exact result.   

\subsection{Model I: the Dephasing model }
Since the exact expression for $L$ is known, the error defined by Eq. \eqref{errorE} is discussed explicitly in this case. Obviously, Error shows strong dependence to the values of $R$ and $N_k$, as illustrated in Figs. \ref{fig:errordephasing}. For a fixed $N_k$, increasing  $R$  leads to a significant reduction in the Error, as shown in panels (a1) and (a2). However, it is observed that the Error starts to fluctuate significantly after a certain time  period, which is obviously influenced by the value $N_k$. This time period, labeled by $t_{p}$, can be compressed when $R$ is increased. Conversely, when $R$ is kept constant, increasing $N_k$ can significantly prolong $t_{p}$, although the error does not display a noticeable reduction as shown by panels (b1) and (b2). Conclusively,  the value $R$ decides the accuracy of evaluation, while $N_k$ is responsible for the efficiency of computation. A simple relationship of $N_k$, $R$ and $t_{p}$ can be established
\be\label{rtp}
R t_{p}\simeq 2 N_k,
\ee
which demonstrates clearly the combined influence of $R$ and $N_k$ on the evaluation of $L$.

Some remarks should be made for Eq. \eqref{rtp}.  A similar relation can be noted in Ref. \cite{vega15}, in which the maximal time $t_{\max}$ of evolution is linearly related to the degree $N_k$ of real orthogonal polynomials, i.e., 
\be \label{vega}
\omega_{\max} t_{\max} = 2 \left(2 N_k+1 \right). 
\ee 
$\omega_{\max}$ is the maximal frequency adopted to approximate the frequency integral in the interval $\left[0, \infty \right)$, which corresponds to $2\omega_c R$ in this paper.  Despite their similar forms, the two relationships have distinct physical meaning. To numerical evaluation of frequency integral, $\omega_{\max}$ is typically on the order of $\sim 100$ at least. However, it is shown in this subsection that very high computational precision can be achieved even for $R$ values as small as $2$. This suggests that our approach may significantly improve computational efficiency compared to the method used in Ref. \cite{vega15}. 
The difference  can be attributed to the distinct discretization strategies employed by the authors of Ref. \cite{vega15}. Specifically, Ref. \cite{vega15} discretizes directly the hybridization function or its time evolution, instead of  introducing unitary transformation to establish the relationship between continuous and discrete operators, as done in this paper and Ref.\cite{discretization1}. In real scenarios, both approaches may reach  the same effective Hamiltonian only when $w(x)$ is chosen as $J(x)$. If not, $H_{\text{int}}$  may become   complicated  requiring further simplification to achieve a compact form, as demonstrated by Eq. \eqref{realhint} in \ref{appendix:realGQ}. Therefore, Eq. \eqref{vega} is a result of special consideration, and cannot be applied for the current discussion.

\begin{figure}[tb]
\center
\includegraphics[width=6cm]{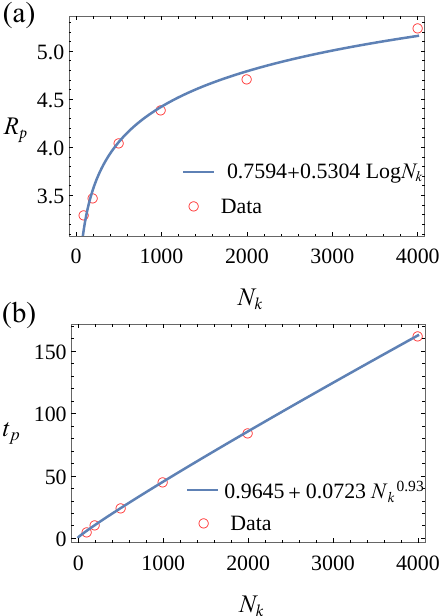}
\caption{(Color online)(a) The logarithmic fit to $R_p$ as a function of $N_k$. (b) The power fit to $t_p$ as a function of $N_k$   }
\label{fig:Rptp}
\end{figure}

\subsection{Model II: the Single-excitation dynamics in AAH model}

Due to the lack of  the exact expression for SP in this model, we conducted a detailed investigation of the Deviation defined by Eq. \eqref{errorA} for $\Delta=1$ in Figs. \ref{fig:errordelta1}. This choice was made because the analytical approach in this scenario can  characterize both the short term  and long term  behavior of SP, leading to clear conclusion. The simulation of SP is carried out using Eq. \eqref{stablepsit} in the plots, with a focus on the region where the modulus of Deviation is less than $10^{-2}$. In Figs. \ref{fig:errordelta1}, the boundary of this region is highlighted by the dashed white line. It is evident that the size of this region displays  non-monotonic variance with the increase of $R$.  Especially, after a enlargement for a relatively small $R$, this region becomes compressed when  $R$ exceeds a special value, termed as $R_p$. However,  the value of Deviation can be significantly reduced when  $R$ is greater than $R_p$.
This finding means that one have to choose properly the value of $R$ to attain the required accuracy.  Obviously, $R_p$ is significantly relevant to the value of $N_k$. The logarithmic fit to $R_p$  is presented as a function of $N_k$ in Fig. \ref{fig:Rptp} (a).

Remarkably, the maximum duration of evolution $t_p$  related to $R_p$ can be found from Fig. \ref{fig:errordelta1},  which  is determined by the intersection of two dashed white lines, if the modulus of Deviation is not exceed 0.01.  Similarly,  $t_p$ signifies the duration during which the simulation remains effective. The relationship of $t_p$ with $N_k$ is shown  in Fig. \ref{fig:Rptp} (b), showing that $t_p$ increases almost linearly with $N_k$.

It is important to note that both $R_p$ and $t_p$ are closely related to the desired threshold of error. A lower Deviation leads to a smaller $t_p$ and a larger $R_p$ for a given $N_k$.  Therefore, a balance must be struck between computational precision and the effective simulation time.  The study for $\Delta=2$ and $\Delta=6$ the similar picture. 

\section{Conclusion}

This paper proposes  a general approach to approximate the environment in continuum using gauss quadratures in the complex frequency space, which is termed as the complex discretization approximation. By this approach, an effective total Hamiltonian can be founded generally, which is nonHermitian and demonstrates complex energy mode with negative imaginary part.  The nonHermitian Hamiltonian is used  to simulate the  dynamics in two exactly solvable models, the dephasing model and single-excitation  dynamics in the AAH model. Compared to the simulation in real frequency spaece, our approache can not only   offer  a significant advantage in compressing the recurrence of the initial-state information, but also  provides a comprehensive  description for the dynamics of open system. 

By the analysis of computational error, our approach demonstrates rapid convergency for calculation when increasing $R$ or $N_k$. This observation means that the calculation by complex discretization approximation is reliable and efficient.  In addition, the error analysis reveals a simple relationship between the parameters in computation and the effectiveness  of calculation. This means that one can manipulate the simulation in a controllable manner. These properties guarantee that even if there is no exact or analytical result,  our approach may provide accurate simulation for the dynamics. 

Finally, it should be stressed that the complex discretization approximation is fundamentally different from the theory of pseudomode \cite{garrway96,tamascelli18,lambert19,xu19,pleasance20}.  In the latter, the influence of bath is characterized by the poles of  spectral function $J(\omega)$ in the complex frequency space. In addition, the poles  are embedded respectively into an individual Markovian environment, of which the dynamics  is captured by the Markovian Lindblad master equation. Moreover, the number of poles is definite for a given $J(\omega)$. In contrast, for the current approach,  the number of complex $\omega$ mode can be arbitrary in this case, depending only on the requirement of computational accuracy.   It is also evident that for Ohm spectral function, there is no pole in the complex $\omega$ plane.  Thus, the  pseudomode method can not be applied directly. 


\section*{ACKNOWLEDGEMENTS}
H.T.C. acknowledges the support of Natural Science Foundation of Shandong Province under Grant No. ZR2021MA036. Y. A. Y. acknowledges the support of National Natural Science Foundation of China (NSFC) under Grant No. 21973036.  M.Q. acknowledges the support of NSFC under Grant No. 11805092 and  Natural Science Foundation of Shandong Province under Grant No. ZR2018PA012. X.X.Y. acknowledges the support of NSFC under Grant No. 12175033 and National Key R$\&$D Program of China (No. 2021YFE0193500).








\appendix

\section{Gauss quadratures  and the mapping to chain Hamiltonian}\label{appendix:realGQ}

This appendix  provides a succinct overview of the Gauss quadratures (GQ) in real frequency space and the process for converting the continuum setting into the chain form. The presentation follows mainly the references \cite{discretization1} and \cite{wilf}.

\emph{Gauss quadratures}
GQ is introduced to improve the numerical integration. Different from the equally spaced abscissas  in the Newton-Cotes formula,  GQ allows the freedom to choose not only the weight coefficients, but also the location of the abscissas at which the function is to be evaluated. It is crucial for GQ  to construct orthogonal polynomials.  For this purpose, one first defines the inner product for any polynomials $f(x), g(x)$
\be\label{inp}
\left\langle f, g \right\rangle=\int_{a}^{b} \text{d}x w(x) f(x) g(x),
\ee
in which $w(x)\geq 0$ is the  weight function defined on interval $x\in \left[a, b\right]$.
Thus the polynomial $p_n(x)=x^n+ \sum_{i=0}^{n-1} A_i x^i$ of degree $n$  is called  orthogonal to  $p_m(x)$ if it satisfies the relation
\be \label{realor}
\left\langle p_m, p_n \right\rangle=\int_{a}^{b} \text{d}x w(x) p_m(x) p_n(x)=0 (m\neq n).
\ee
With the assumption  $p_0(x)=1$ and  $p_{-1}(x)=0$, the recurrence relation can be deduce by Eq. \eqref{realor},
\be\label{precurrence}
p_{n+1}(x)= \left(x - \alpha_n\right)p_n(x)- \beta_n p_{n-1}(x),
\ee
in which
\be
\alpha_n=\frac{\left\langle x p_n,p_n\right\rangle}{\left\langle p_n, p_n\right\rangle}, \beta_n=\frac{\left\langle p_n, p_n\right\rangle}{\left\langle p_{n-1}, p_{n-1}\right\rangle}.
\ee
The proof can be found in Ref. \cite{discretization1} or \cite{wilf}.

The determination of the  zero points of $p_n(x)$, denoted as $x_i (i=1,2, \cdots, n)$, are crucial for the numerical integration.  In order to find $x_i$, it is convenient to introduce orthonormalized  polynomials $\pi_n(x)=p_n(x)/\sqrt{\left\langle p_n, p_n\right\rangle}$. 
Thus one has the orthonormality relation
\be \label{nor}
\left\langle \pi_m, \pi_n \right\rangle=\int_{a}^{b} \text{d}x w(x) \pi_m(x) \pi_n(x)=\delta_{m, n}.
\ee
The recurrence relation becomes
\be \label{nrecurrence}
\sqrt{\beta_{n+1}} \pi_{n+1}(x)= \left(x - \alpha_n\right) \pi_{n}(x) -\sqrt{\beta_{n}}\pi_{n-1}(x).
\ee
Alternatively, Eq. \eqref{nrecurrence} can be rewritten in a matrix form
\be\label{Mrecurrence}
&&x \left( \begin{array}{c}
             \pi_0 \\ \pi_1 \\ \vdots \\ \pi_{n-1}
           \end{array}\right)
= M_r
\left( \begin{array}{c}
             \pi_0 \\ \pi_1 \\ \vdots \\ \pi_{n-1}
           \end{array}\right)\nonumber
+ \sqrt{\beta_n}\left( \begin{array}{c}
             0\\ 0 \\ \vdots \\ \pi_{n}\end{array}\right); \\
&&M_r=\left(\begin{array}{cccc}
\alpha_0 & \sqrt{\beta_1} & 0 & \cdots \\
\sqrt{\beta_1} & \alpha_1 & \ddots & 0 \\
0 & \ddots &  \ddots & \sqrt{\beta_{n-1}} \\
0 & \cdots & \sqrt{\beta_{n-1}} & \alpha_{n-1}
   \end{array}\right),
\ee
in which $\pi_n$ implies $\pi_n(x)$. Thus, $x_i$ corresponds to the  eigenvalue  of symmetric matrix $M_r$. The corresponding orthonormalized eigenvector $\ket{x_i}$ can be determined by normalizing  $\left(\pi_0 (x_i), \pi_1(x_i), \cdots, \pi_{n-1}(x_i)\right)^T$.

The weight $w_i$ for  $x_i$ can be decided by the Christoffel-Darboux identity. It has been proved that  $w_i$  satisfies the relation \cite{wilf}
\be
w_i \sum_{k=0}^{n-1} \left[ \pi_k(x_i)  \right]^2 =1  \left( i=1, 2, \cdots, n \right).
\ee
In practice,  it is more convenient to determine   $w_i$ by the equivalence
\be
\sqrt{w_i} \left(\pi_0 (x_i), \pi_1(x_i), \cdots, \pi_{n-1}(x_i)\right)^T \Rightarrow \ket{x_i}.
\ee
By $\ket{x_i}=\left(q^{(i)}_0, q^{(i)}_1, \cdots, q^{(i)}_{n-1} \right)^T$, which can be obtained directly by numerics, one gets
\be \label{wi}
w_i=\left[\frac{q^{(i)}_0}{  \pi_0(x_i)}\right]^2=\left[q^{(i)}_0\right]^2 \int_{a}^{b} \text{d}x w(x).
\ee

In summary, the following theorem can be found \cite{wilf},
\begin{theorem}\label{GQtheorem}
  Let $w(x)$ be a weight function on the internal $\left[a, b\right]$. Then the polynomials $\pi_n(x)$ can present real zero points  $x_i$ and corresponding weight $w_i (i= 1, 2, \cdots, n)$, which show the properties (i) $a<x_i<b (\forall i)$; (ii) $w_i > 0 (\forall i)$; (iii) The equivalence
  \be
  \int_{a}^{b} \text{d}x w(x) f(x) = \sum_{i=0}^{n-1} w_i f(x_i)
  \ee
  is exactly true for polynomial $f(x)$ of degree $\leq (2n-1)$.
\end{theorem}
Some comments should be made in this place.  First, the choice of $w(x)$ is important, as it determines entirely the polynomials $\pi_n(x)$. Generally, there is no mathematical restriction on the choice of $w(x)$. However, it is ideal to relate $w(x)$ with the physical quantity. For instance it can be constructed, based on the spectral function $J(x)$, as shown in the following illustrations. Secondly, $x_i$ displays different contribution to the evaluation, weighted by $w_i$. In regard to the open quantum system, it implies that certain modes of environment would dominate the open dynamics of system. Finally, in the event that $f(x)$ is a series or the upper bound  approach infinity, the integration can solely be deemed precise only if $n$ approaches infinity. Thus, the degree of accuracy in the evaluation is contingent upon the value of $n$.

\emph{Mapping to Chain form-} In order to discretize the continuum in environment, one can introduce the transformation \cite{discretization1}
\be\label{dn}
d_n= \int_{a}^{b} \text{d}x  \sqrt{w(x)} \pi_n(x) b_x.
\ee
and the inverse
\be \label{inversedn}
b_x= \sqrt{w(x)} \sum_{n=0}^{N_k-1} \pi_n(x)d_n,
\ee
in which $N_k$ denotes the  degree of polynomial used in evaluation. It can be proved directly
\be
\left[d_n, d_m^{\dagger}\right]&=&\iint \text{d}x \text{d}x' \sqrt{w(x)w(x')} \pi_n(x) \pi_m(x') \left[b_x, b^{\dagger}_{x'}\right]\nonumber \\
&=& \int_a^b \text{d}x w(x) \pi_n(x) \pi_m(x) =\delta_{n, m},
\ee
where $ \left[b_x, b^{\dagger}_{x'}\right]=\delta(x-x')$ is applied.
Substituting Eq. \eqref{inversedn} into  $H_b$, one gets
\be
H_b=\omega_c \sum_{m,n} d_m^{\dagger} d_n \int_{a}^{b} \text{d}x w(x) \pi_m(x) x \pi_n(x),
\ee
in which $\omega(x) = \omega_c x$ is used. Replacing $x \pi_n(x)$ according to  Eq. \eqref{nrecurrence}, one gets
\be\label{discreteHb}
H_b&=&\omega_c\sum_{m, n}\left(\alpha_n \delta_{m, n} + \sqrt{\beta_{n+1}}\delta_{m, n+1}+ \sqrt{\beta_{n}}\delta_{m, n-1}\right) d_m^{\dagger}d_n\nonumber \\
&=& \omega_c \left(\cdots,  d_m^{\dagger}, \cdots\right)M_r
\left( \begin{array}{c}\vdots \\ d_n \\ \vdots \end{array}\right),
\ee
where $M_r$ is given by Eq. \eqref{Mrecurrence}. Evidently, $H_b$ is transformed into a chain form with the nearest neighbor hopping.

Accordingly, $H_{\text{int}}$  can be rewritten as
\be
H_{\text{int}}= \sum_{n}\sum_{i=1}^{N_k}\int_a^{b} \text{d}x \sqrt{w(x)} \pi_i(x)\left[g(x)c^{\dagger}_n d_i + g^*(x)d_i^{\dagger} c_n\right]. \nonumber \\
\ee
Assuming that $g(x)$ is real and setting $w(x)=g^2(x)$, one thus obtain
\be \label{discreteHint}
H_{\text{int}}&=& \sum_{n}\sum_{i=1}^{N_k}\int_a^{b} \text{d}x w(x) \pi_i(x)\left[c^{\dagger}_n d_i + d_i^{\dagger} c_n\right] \nonumber\\
&=& \sqrt{\int_{a}^{b} \text{d}x w(x)}\sum_{n} \left(c^{\dagger}_n d_0 + d_0^{\dagger} c_n\right),
\ee
in which the orthonormality relation Eq. \eqref{nor} is applied for the second equality. It is evident that Eq. \eqref{discreteHint} features the coupling of system to the end of chain, depicted by Eq. \eqref{discreteHb}.

Eqs. \eqref{discreteHb} and \eqref{discreteHint} are subject to certain comments. First,  the transformation Eq. \eqref{inversedn} is unitary only when $N_k \rightarrow \infty$. Only in this situation, Eqs. \eqref{discreteHb} and \eqref{discreteHint} can be considered equivalent exactly to $H_b$ and $H_{\text{int}}$ in Eq. \eqref{H}. Therefor, a large $N_k$  as possible is necessary to  accurately simulate the open dynamics of system. However, it makes the computation  very exhaustive. Secondly, a reasonable choice of $w(x)$ can facilitate numerical simulation. Evidently, the choice of $w(x)=g^2(x)$ gives the  straightforward physical interpretation for  $H_{\text{int}}$.  Finally, by diagonalizing  Eq. \eqref{discreteHb}, one gets
\be\label{dhb}
H_b&=&\omega_c \sum_{i=1}^{N_k} x_i \tilde{d}_i^{\dagger}  \tilde{d}_i \\
\label{dhint}
H_ {\text{int}}&=&\sum_{n}\sum_{i=1}^{N_k} \sqrt{w_i} \left(c^{\dagger}_n \tilde{d}_i  + \tilde{d}_i^{\dagger}  c_n\right)
\ee
where $\tilde{d}_i^{(\dagger)}= \sqrt{w_i} \sum_n \pi_n(x_i) d^{(\dagger)}_n$ denote the  annihilation (creation) operator relevant to the energy level $\omega_c x_i$. It is obvious that $x_i$ characterizes the discrete energy mode in environment. As a consequence, the discretization approximation for environment is reduced to find the zero points of polynomial $\pi_{N_k}(x)$.

However, for arbitrary $w(x)\neq g^2(x)$, one notes
\be 
H_{\text{int}}= \sum_{n}\sum_{i=1}^{N_k}\int_a^{b} \text{d}x w(x)  \frac{\pi_i(x)}{\sqrt{w(x)}}\left[g(x)c^{\dagger}_n d_i + g^*(x)d_i^{\dagger} c_n\right].\nonumber 
\ee
By Theorem \ref{GQtheorem}, 
\be 
\int_a^{b} \text{d}x  \frac{w(x) \pi_i(x)}{\sqrt{w(x)}} g^{(*)}(x)\simeq \sum_j w_j\frac{g^{(*)}(x_j)}{\sqrt{w(x_j)}} \pi_i(x_j)
\ee
Substituting the relationship above into $H_{\text{int}}$, one gets
\be \label{realhint}
H_{\text{int}}&=& \sum_{n} \sum_{j=1}^{N_k}\left[ c^{\dagger}_n  \sqrt{w_j} \frac{g(x_j)}{\sqrt{w(x_j)}}  \left( \sum_{i=1}^{N_k} \pi_i(x_j)\sqrt{w_j}   d_i \right) + \text{h. c.} \right]  \nonumber \\
&=& \sum_{n} \sum_{j=1}^{N_k}\left[ \sqrt{w_j} \frac{g(x_j)}{\sqrt{w(x_j)}} c^{\dagger}_n \tilde{d}_j   + \text{h. c.} \right], 
\ee
where $\tilde{d}_j^{(\dagger)}= \sqrt{w_j} \sum_i \pi_i(x_j) d^{(\dagger)}_i$ is used for second equality. The validity of this approach has been checked explicitly by comparison to the exact results.

\begin{figure*}[tb]
\center
\includegraphics[width=17cm]{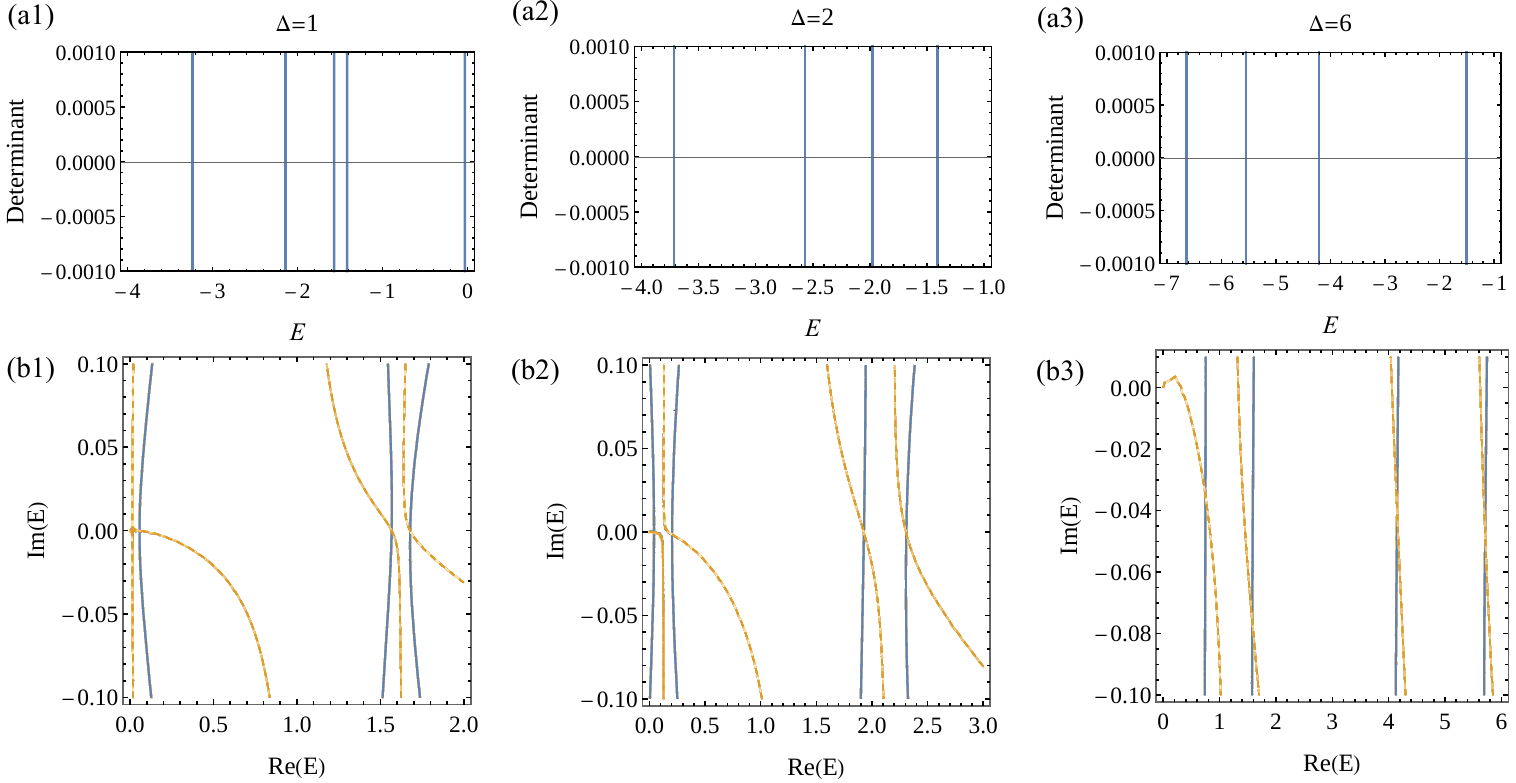}
\caption{(Color online) (a1) - (a3): The plots for the determinate of coefficient matrix of  Eq. \eqref{bs} when $E<0$. (b1) -(b3): The  determinate of coefficient matrix of  Eq. \eqref{bs} is plotted by a contour of value zero for the complex $E$ with positive real part. The solid line denotes the real part of the determinant, and the dashed line denotes the imaginary part. The intersection implies the existence of the decaying energy level. For all plots, we have chosen $N_s=8,  \beta=\left(\sqrt{5}-1\right)/2, \phi=\pi $ and $\eta=0.1, \omega_c=10$. The  $E$ is in units of $J$}
\label{fig:bs}
\end{figure*}

\section{Analytical approach to the single excitation open dynamics in AAH model }\label{analyticalsolution}

This appendix provides the analytical approach to the single-excitation open dynamics of AAH model by solving  Eq. \eqref{Ap}. Mathematically, Eq. \eqref{Ap} represents the  linear system of equation for the unknown $A_n(p)$s. In principle, one can find all $A_n(p)$ by Cramer's rule, only if the integral $ \int_{0}^{\infty} \text{d}\omega \frac{J(\omega) }{ \omega- \mathbbm{i} p}$ is evaluated accurately. Then, $a(t)$ can be obtained analytically using inverse Laplace transformation. There are two specific scenarios, where the solution to Eq. \eqref{Ap} can analytically determined.

\emph{-The single particle bound state-}
When $\mathbbm{i}p= E $ with  $E<0$, the physical interpretation of $E$ can be elucidated through inverse Laplace transformation, defined formally as
\be\label{inverselaplace}
a_n(t)=\frac{1}{2\pi \mathbbm{i}} \int_{s-\mathbbm{i}\infty}^{s+\mathbbm{i}\infty} \text{d}p A_n(p) e^{p t}.
\ee
In analogy to the evolution operator in quantum mechanics, $A_n(p)$ can be viewed as the probability amplitude, with $E$ representing the eigenvalue of the Hamiltonian. This relationship between $E$ and the corresponding $A_n(p)$ defines the unitary evolution that governs the stable behavior in the open dynamics of the system.

Typically, the value of $E$ is determined by solving the equation $H \ket{\psi}= E \ket{\psi}$. In the single-excitation subspace, the total state  $\ket{\psi}$ can be written formally as
\be
\ket{\psi}= \left(\sum_{n=1}^{N_s} a_n \ket{1}_n \right) \ket{0}^{\otimes k} +\ket{0}^{\otimes N_s}  \left(\sum_{k} b_k \ket{1}_k\right).
\ee
Eliminating the dependence on $b_k$, one gets
\be\label{bs}
a_{n+1}  &+& a_{n-1} + \Delta \cos(2\pi \beta n +\phi)a_n  +\nonumber \\
&& \sum_{n=1}^N a_n \int_0^{\infty} \text{d}\omega\frac{J(\omega)}{E-\omega} = E a_n.
\ee
The solution to the above equation in the region $E<0$ can be determined by setting the determinant of the  coefficient matrix for variables $\left(a_1, a_2, \cdots, a_{N_s} \right) ^T$ equal to zero. In Figs. \ref{fig:bs}(a1)-(a3), the determinant of  coefficient matrix is plotted  to identify  the values of $E$. Evidently, the intersection of the determinant and the $E$-axis can be found at some isolated values of $E$. These discrete  energy levels are denoted  as the single particle bound state s \cite{kofman1994}.  Because of the finite energy gap from the continuum $\omega_k \in [0, \infty)$, the single particle bound state can display the robustness against the dissipation induced by the coupling to environment \cite{tong10,xiong10,tong11}.

Once $p = - \mathbbm{i}E_k$ is given, the values of  $\left( A_1(p), A_2(p), \cdots, A_{N_s}(p) \right)$ can be derived by solving Eq. \eqref{Ap}. Since the coefficient matrix in Eq. \eqref{Ap} is identical to that in Eq. \eqref{bs},  $E_k$ represents a singular point in the  expression for $A_n(p) \left(n=1, 2, \cdots, N_s \right)$, which can be obtained by applying Cramer's rule to  Eq. \eqref{Ap}. This singularity allows for the straightforward evaluation of  Eq. \eqref{inverselaplace} using the residual theorem, i.e.
\be
a_n(t) &\rightarrow& \sum_k c_{n}(E_k) e^{- \mathbbm{i}E_k t}, \\
 c_{n}(E_k) &=&\lim_{E\rightarrow E_k}  A_n(p) \left(E - E_k\right). \nonumber
\ee

\emph{-The decaying state-}
The integral $ \int_{0}^{\infty} \text{d}\omega \frac{J(\omega) }{ \omega- E}$ becomes divergent when $E>0$. This problem can be resolved by adding an  imaginary item in the denominator, such as  $\omega \rightarrow \omega - \mathbbm{i}\epsilon $  with infinitesimal $\epsilon>0$. Then by Sokhotski-Plemdj (SP) formula
\be
\lim_{\epsilon\rightarrow 0} \frac{1}{x- x_0 - \mathbbm{i}\epsilon} = \text{P} \frac{1}{x -x_0}  + \mathbbm{i} \pi \delta \left( x- x_0\right),
\ee
one can get
\be
\lim_{\epsilon\rightarrow 0} \int_{0}^{\infty} \text{d}\omega \frac{J(\omega) }{ \omega- E- \mathbbm{i}\epsilon} = \text{P} \int_{0}^{\infty} \text{d}\omega \frac{J(\omega) }{ \omega- E} + \mathbbm{i} \pi J(E), \nonumber \\
\ee
where $\text{P}$ denotes the principle value. In this manner, the decaying energy level can be determined. The decaying energy level is complex and displays a negative imaginary part, which can characterize the dissipation in the system.

Assuming  $E= x + \mathbbm{i}y$ with real $x, y$ and $x>0$, and substituting this expression  into Eq. \eqref{bs}, one can get the decaying energy level  by setting the determinant of  coefficient matrix equal to zero. In Fig. \ref{fig:bs} (b1)-(b3), the real and imaginary parts of the determinate is plotted by a contour  of zero value, in which the intersection of the solid and dashed line represents the occurrence of the decaying energy level. The explicit calculation shows that the decaying energy level closed to the axis $\text{Im}(E)=0$ has negative imaginary part, which accurately describes  the open dynamics in AAH model. The corresponding inverse Laplace transformation can be carried out in the same way as described in the previous part. However, it is noted that the intersection can also be found far away from the axis $\text{Im}(E)=0$. The explicit calculation shows that these state would be unphysical since they can introduce the gain effect in the dynamics or result in a  SP  greater than unity.  Unfortunately, we do not know currently  how to preclude the unphysical intersection during  the evaluation.

Conclusively, one can determine the open dynamics of AAH model analytically using
\be\label{ant}
a_n(t)\simeq \sum_k c_{n}(E_k) e^{- \mathbbm{i}E_k t} + \sum_k c_{n}(x_k+  \mathbbm{i} y_k) e^{- \mathbbm{i}\left( x_k + \mathbbm{i} y_k \right) t} \nonumber \\
\ee
The first summation in the equation above representing the contribution of the single particle bound state, depicts the long term behavior of the dynamics, while the second summation characterizes the intermediated term dynamics from the decaying states. As shown in Fig. \ref{fig:aah}, the calculation  by this equation  aligns well with the exact numerical evaluation and the simulation.  However, for a very small $t$, Eq. \eqref{ant} may not be accurate and  the exact numerical evaluation is recommended in such case.

\end{document}